\documentclass[aps,prl,twocolumn,showpacs,superscriptaddress]{revtex4-1}  % for review and submission

\usepackage{graphicx}  % needed for figures
\usepackage{dcolumn}   % needed for some tables
\usepackage{bm}        % for math
\usepackage{amssymb}   % for math
\usepackage{amsmath}   % for math
\usepackage{color}

\newcommand{\seq}{$\langle S \rangle_{\mathrm{eq}}$}

\begin{document}

\title{Unifying framework for strong and fragile liquids via machine learning: a study of liquid silica}
\author{Ekin D. Cubuk}
\email{cubuk@google.com}
\affiliation{Google Research, Brain Team}
\author{Andrea J. Liu}
\affiliation{Department of Physics, University of Pennsylvania, Philadelphia, Pennsylvania 19104, USA}
\author{Efthimios Kaxiras}
\affiliation{Department of Physics and School of Engineering and Applied Sciences, Harvard University, Cambridge, Massachusetts 02138, USA}
\author{Samuel S. Schoenholz}
\email{schsam@google.com}
\affiliation{Google Research, Brain Team}
\date{\today}

\begin{abstract}
The fragility of a glassforming liquid characterizes how rapidly its relaxation dynamics slow down with cooling. The viscosity of strong liquids follows an Arrhenius law with a temperature-independent barrier height to rearrangements responsible for relaxation, whereas fragile liquids experience a much faster increase in their dynamics, suggesting a barrier height that increases with decreasing temperature. Strong glassformers are typically network glasses, while fragile glassformers are typically molecular or hard-sphere-like. As a result of these differences at the microscopic level, strong and fragile glassformers are usually treated separately from a theoretical point of view. Silica is the archetypal strong glassformer at low temperatures, but also exhibits a mysterious strong-to-fragile crossover at higher temperatures. Here we show that softness, a structure-based machine learned parameter that has previously been applied to fragile glassformers provides a useful description of model liquid silica in the strong and fragile regimes, and through the strong-to-fragile crossover. Just as for fragile glassformers, the relationship between softness and dynamics is invariant and Arrhenius in all regimes, but the average softness changes with temperature. The strong-to-fragile crossover in silica is not due to a sudden, qualitative change in structure, but can be explained by a simple Arrhenius form with a continuously and linearly changing local structure. Our results unify the study of liquid silica under a single simple conceptual picture.         
\end{abstract}


\maketitle

When cooled quickly enough, liquids can avoid crystallization and relax more and more slowly, eventually undergoing dynamical arrest on human time scales even as their structure remains similar to that of the liquid. In 1985, Angell suggested placing glassforming liquids into a spectrum from strong liquids to fragile liquids in the celebrated ``Angell plot"~\cite{angell1985strong,angell1991relaxation,salmon2013identifying}. In strong liquids like silica, the dynamics of the liquid slow down as one would expect from an Arrhenius process. Most other liquids are fragile, however, and exhibit a much faster slowing of the dynamics. To complicate matters further, silica undergoes a strong-to-fragile crossover upon heating at around 3100-3300 K~\cite{saika2001fragile, geske2016fragile}.  

It has been shown that the non-Arrhenius behavior of a model fragile glassformer, a binary Lennard-Jones mixture known as the Kob-Andersen model~\cite{kob94}, can be captured by simple expression involving a structural quantity known as ``softness" ~\cite{schoenholz2016structural}. This quantity is obtained using machine learning to identify the linear combination of a set of structural quantities that correlates most strongly with rearrangements in the supercooled liquid. At any given moment, different particles have different softnesses described by a distribution. For three fragile glassformers, the Kob-Andersen model~\cite{schoenholz2017relationship}, the Weeks-Chandler-Andersen model~\cite{landes2019attractive} and a model polymer glassformer~\cite{sussman2017disconnecting}, the relaxation time $\tau_\alpha$ can be expressed simply as~\cite{schoenholz2017relationship}
\begin{equation}
\tau_\alpha \propto 1/P_R(\langle S \rangle), \label{eq:relax}
\end{equation}
where $P_R(S)$ is the probability density that a particle of softness $S$ will rearrange and $\langle S \rangle$ is the average softness of particles in the system. The rearrangement probability $P_R(S)$ is Arrhenius at each value of $S$, implying that particles of softness $S$ have a well-defined, temperature-independent free energy barrier to rearrangement $\Delta F(S)$. The barrier increases with decreasing $S$, so that particles with lower softness have a smaller propensity to rearrange. As the system cools, $\langle S \rangle$ decreases; this leads to slowing down of the relaxation time.

% Previous work has shown that a parameter called softness can be used to predict particle rearrangements in a variety of systems, including supercooled Lennard-Jones liquid simulated in computers to colloid and granular pillar experiments in labs. Softness is a description of the local structure of particles, extracted from molecular dynamics (MD) simulations or videos of stress simulations using machine learning.   

In this Letter, we show that the softness-based description of the dynamics outlined above for fragile glassformers also applies to a model of a strong glassformer, namely silica. Further, we show that this description can predict the dynamics not only of the low-temperature strong liquid, but also  the strong-to-fragile crossover and the high temperature fragile liquid. %The slower decay of average softness in fragile liquids at lower temperatures gives rise to the distinct behavior of fragile and strong liquids. 
Thus, our results suggest that the wide diversity of fragility observed in different liquids can be boiled down to the temperature dependence of a single machine-learned variable.   

We model silica liquid using the potential of van Beest, Krame, and van Santen (BKS)~\cite{vanbeest90force}, which has been commonly used to study liquid~\cite{saika2001fragile,horbach1999static} and amorphous silica~\cite{vollmayr1996investigating}. A harmonic potential is used at small distances to prevent Si and O atoms from fusing together~\cite{vollmayr1996investigating}. We confirm that this modification reliably prevents unphysical fusion events without affecting the potential at relevant temperatures. We use a unit cell of 2880 atoms: 960 Si and 1920 O atoms. Simulations are done using the LAMMPS package~\cite{plimpton95,thompson2009general,brown2011implementing,brown2012implementing}. We start by melting an $\alpha$-quartz structure at 6000 K. We use a time step of 1 fs. We then quench the system with a cooling rate of $5\times 10^{12} $ K s$^{-1}$ to the final temperature of 2500 K, using the NPT ensemble at zero pressure. We then fix the density of the system to its value at 2500 K, and switch to the NVT ensemble (for all temperatures studied). Training data is collected at 2500 K, and this single trained model is used throughout the paper across the full range of temperatures. We output states every 400 fs, and quench them to their inherent structure using a combination of FIRE~\cite{bitzek06} and conjugate gradient algorithms. 

The calculation of the softness variable has been described in previous work, in the context of dynamical heterogeneities~\cite{schoenholz2016structural}, plasticity~\cite{cubuk2017structure}, thin film dynamics ~\cite{sussman2017disconnecting}, and grain boundaries in polycrystals~\cite{sharp2018machine}. Here we summarize it for completeness, but readers should refer to previous work for further details. For each atom in the quenched state, we calculate the quantity $p_{hop}$ (see Methods for details), which (with a time window of 4ps) labels atoms as either rearranging (\textit{i.~e.~} about to rearrange in the next time window) or non-rearranging (\textit{i~.e~.} not going to rearrange for a time of order the relaxation time). We then parameterize the local structure around each atom using a set of structure functions~\cite{cubuk15, schoenholz2016structural} which are inspired by and very similar to widely-used symmetry functions~\cite{behler07,NN_DFT_diamond,behler2015constructing,artrith2011high,artrith2012high,artrith2013neural,artrith2016implementation,cubuk2017representations, onat2018implanted}. Given a set of structure functions, the local structural environment of an individual atom, $i$, can then be described by a point in structure-function space.

We then use a support vector machine (SVM)~\cite{SVM,libsvm,fan2008liblinear} to train a classifier to distinguish between a set of rearranging and non-rearranging atoms (a different SVM is trained for Si and O atoms), based on their local structure. Training the SVM leads to a classification hyperplane with particles on one side being classified as not susceptible to rearrangement, while particles on the other side are likely to rearrange.  The test-set accuracy of this model is found to be 86\%, which is slightly lower than the 90\% accuracy that was achieved on the simpler system of Lennard-Jones particles~\cite{schoenholz2016structural}. An interesting observation is that although restricting the model to only use radial structure functions leads to only a 2\% decrease for Lennard-Jones systems~\cite{schoenholz2016structural}, it leads to a significantly larger 7\% decrease for silica. This is not surprising given the directional bonding present in SiO$_2$ compared with the spherically-symmetric interactions observed in the Lennard-Jones model. A similar conclusion was reached regarding the importance of three-body interactions in modeling silica dynamics~\cite{kob2002quantitative}. 

The softness, $S_i$, of atom $i$ is defined as the signed distance between that point and the classification hyperplane.  Atoms on the rearranging side of the hyperplane have $S_i>0$, whereas atoms on the non-rearranging side have $S_i<0$. Previous analysis on a variety of systems with isotropic interactions has found that not only does the probability of rearranging, $P_R(S)$, have the Arrhenius form, $P_R(S) \sim \exp [\Delta F(S)/T]$ with temperature $T$, but that the free energy barrier to rearrangement, $\Delta F(S)$  decreases approximately linearly with increasing $S$~\cite{schoenholz2016structural, schoenholz2017relationship, sussman2017disconnecting, sharp2018machine, freitas2020uncovering, landes2019attractive}.

\begin{figure}
\includegraphics[width=0.5\textwidth]{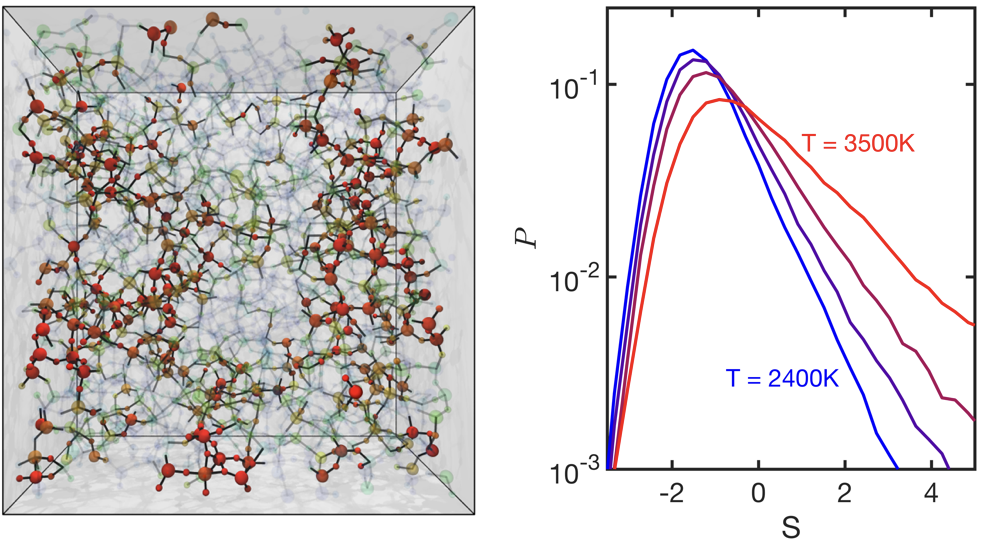}
\caption{
(a) A snapshot of the SiO$_2$ unit cell at 2500 K. Softness is represented on an opacity scale with $S \ll 0$ being transparent and $S 
\gg 0$ being opaque. (b) Distribution of softness, $P(S)$ at temperatures 2400 K, 2700 K, 3000 K, and 3500 K, with line types ranging from blue to red.
}
\label{Fig1}
\centering
\end{figure}

To study liquid silica in the strong and fragile regimes, we run MD simulations at several temperatures between 2400 K and 6000 K. Following previous work, we train the SVM at data collected at a low temperature (2500 K), and apply this SVM at all the other temperatures. In Fig.~\ref{Fig1}a, we show a representative structure of the SiO$_2$ unit cell at 2500 K. Fig.~\ref{Fig1}b shows the softness distribution at several temperatures. Note that for the fragile glassformers studied previously, the distribution of softnesses is well approximated by a Gaussian distribution - that is slightly skewed towards high softness - with a mean that increases with temperature and the standard deviation that remains roughly constant~\cite{schoenholz2016structural}. Fig.~\ref{Fig1}(b) shows that the softness distributions are very different for silica. They are non-symmetric at all temperatures studied, with a growing high-$S$-tail with increasing temperature.  The softness distribution of silica atoms appears to be well-approximated by the Gumbel distribution~\cite{gumbel1935valeurs}, which characterizes the extreme values of a number of samples from a distribution. An intriguing possibility is that the Gumbel distribution arises because for each atom $i$ the barrier $\Delta F(S_i)$ is roughly equal to the smallest barrier accessible to that atom. However, it is unclear why the same argument would not apply to the fragile Lennard-Jones systems. This issue would be interesting to investigate in future work.    

\begin{figure}
\includegraphics[width=0.5\textwidth]{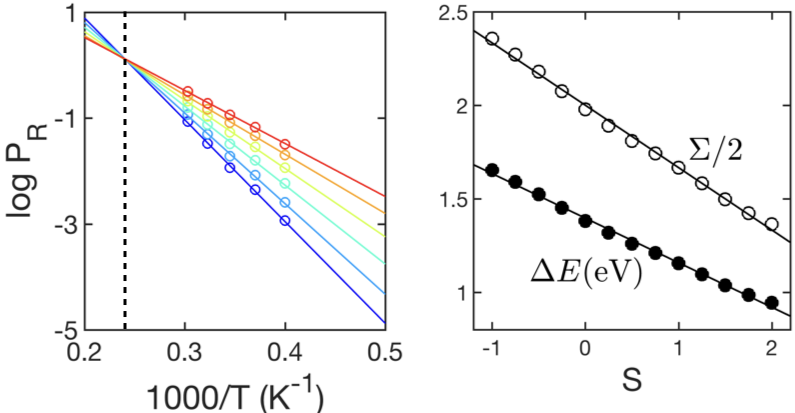}
\caption{
(a) Probabilities of rearrangement as a function of inverse temperature for six different softness values from $S\sim -1$ (blue) to $S\sim 2.75$ (red). Circles denote measurements from MD, solid lines are the corresponding Arrhenius fits. Vertical dashed line indicates the onset temperature ($\sim$ 4150 K). (b) Extracted energy barrier and prefactor from the Arrhenius fits in (a). 
}
\label{Fig2}
\centering
\end{figure}

We examine the probability of rearrangement $P_R(S)$, given by the fraction of particles of softness $S$ that are rearranging at a given time step, averaged over time steps. The temperature-dependence of $P_R(S)$ at each $S$ was found to be Arrhenius for particles with spherically-symmetric interactions~\cite{schoenholz2016structural,schoenholz2017relationship}, for polymers made up of monomers with spherically-symmetric interactions plus anisotropic bonding along the polymer backbone~\cite{sussman2017disconnecting} and aluminum atoms in polycrystals~\cite{sharp2018machine}). We find this to be true for SiO$_2$ as well (Fig.~\ref{Fig2}a), implying that the rearrangements in silica are governed by Arrhenius processes where the free energy barrier to rearrangement is decided by the local structure of the atoms (characterized by $S$). . This is surprising since SiO$_2$ behaves significantly differently at low and high temperatures. We fit $P_R(S)$ to the Arrhenius form:
\begin{equation}
P_R(S) = \mathrm{exp}(\Sigma(S)-\Delta E(S)/T),
\label{eq:arrhenius}
\end{equation}
where $\Sigma(S)$ and $\Delta E(S)$ are fitting constants for each $S$ and the free energy barrier is interpreted as $\Delta F(S) \equiv E(S)-T\Sigma (S)$. We plot $\Delta E(S)$, the energy barrier to rearrangement, and $\Sigma(S)$, the temperature-independent term, as a function of $S$ in Fig~\ref{Fig2}b. 

\begin{figure}
\includegraphics[width=\linewidth]{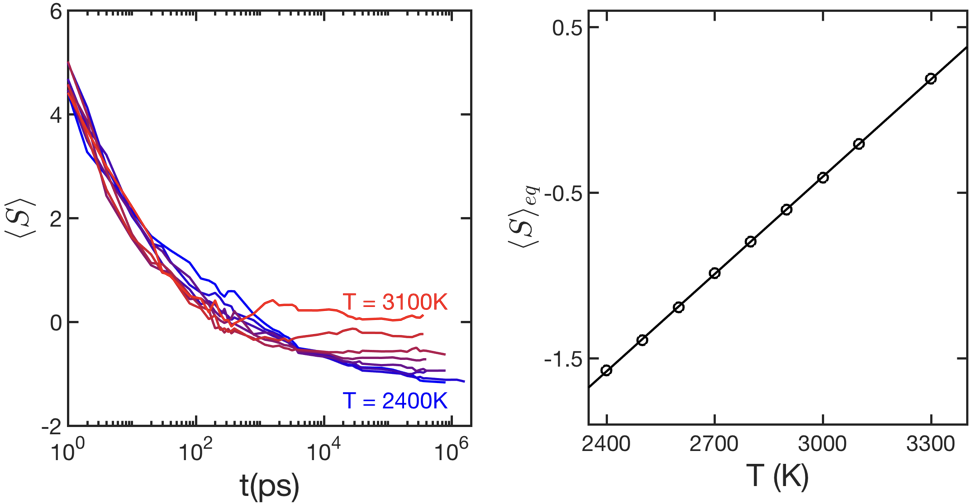}
\caption{
(a) Average softness as a function of waiting time at temperatures $T=2400$ K to 3100 K. The line-type/color scheme is the same as in Fig.~1. (b) Equilibrium softness values vs.~T. The line is a fit to the equilibrium softness values.
}
\label{Fig3}
\centering
\end{figure}

Given that the dynamics are Arrhenius at all temperatures, what causes the strong-to-fragile crossover as $T$ increases?  To answer this question, we study how $S$ itself changes as a function of time. We begin by equilibrating the silica melt at 6000 K, and then rapidly lower the temperature to $T_f$. We then follow the system as it approaches equilibrium at fixed temperature and volume. In Fig~\ref{Fig3}(a), we show how the average softness of particles, $\langle S \rangle$ changes as a function of waiting time, $t_w$, following the temperature quench for several temperatures between 2400 K and 3100 K. It is interesting to note that the average softness, $\langle S \rangle$, of systems with different $T_f$ decays at the same rate until a system equilibrates; this was also observed for the Kob-Andersen glassformer~\cite{schoenholz2017relationship}. For systems that equilibrate within $10^6$ ps, we estimate \seq by direct measurement. For systems that do not equilibrate within our simulation timescale, we investigate the functional form of $\langle S \rangle(t_w)$. We find that it can approximated by a power law function of the form $\langle S \rangle(t_w) = C \cdot t_w^{-\alpha} + \langle S \rangle_{\mathrm{eq}}$. By curve-fitting to the data shown in Fig.~\ref{Fig3}(a), we extrapolate to obtain $\langle S \rangle_{\mathrm{eq}}$ as a function of temperature. The results are shown in Fig.~\ref{Fig3}(b).        

The linear fit to $\langle S \rangle_{\mathrm{eq}}(T)$ is strikingly good. It shows that the average scale for the energy barriers encountered in liquid silica is linear in temperature, at least in the range that we have studied (2400 K - 6000 K). This linear behavior can be contrasted with the phenomenology of silica. At the low temperatures near 2400K, liquid silica behaves as a strong liquid, where the dynamics scale with temperature in an Arrhenius fashion~\cite{bohmer1993nonexponential}. As the temperature is increased, silica transitions into a fragile liquid~\cite{saika2001fragile}. 

To explore fragility vs.~$T$ in terms of softness, we must relate  $\langle S \rangle_{\mathrm{eq}}(T)$ to relaxation. In silica, relaxation is typically quantified by the diffusion constant, defined via the long-time behavior of the mean squared displacement,
\begin{equation}
    \lim_{t\to\infty}\langle (r(t) - r(0))^2\rangle \sim D t.
\end{equation}
We now follow the considerable empirical evidence and assume that particles diffuse via discrete hops that occur intermittently compared with caged vibrations. We take the timescale for vibrations to be $\tau_\mathrm{vib}\approx 400$~fs. Thus, in a similar spirit to~\cite{niblett2016dynamics}, we write
\begin{equation}\label{eq:displacement}
    r(t) = \sum_{n=0}^{t/\tau_\mathrm{vib}} \Delta r_n + r(0)
\end{equation}
where $\Delta r_n$ is the displacement during a hop that occurs (or does not occur) at a time $n\tau$. 

To analyze Eq.~\eqref{eq:displacement}, we make the additional approximation that hops are independent in time such that $\langle \Delta r_i\Delta r_j\rangle = \langle(\Delta r_i)^2\rangle \delta_{ij}.$ While this does not hold exactly - especially when $i = j \pm 1$ - for events that do not occur in succession this approximation is reasonable~\cite{niblett2016dynamics}. Moreover, we assume the system is in equilibrium so that $\langle(\Delta r_i)^2\rangle = \langle(\Delta r_0)^2\rangle = \langle(\Delta r)^2\rangle$. It follows that the diffusion constant is related to the hop statistics by
\begin{equation}\label{eq:diffusion_raw}
    D \sim \langle (\Delta r)^2\rangle.
\end{equation}
As we have hinted at above, Eq.~\eqref{eq:diffusion_raw} is difficult to analyze since $\Delta r$ contains contributions from both hopping and non-hopping particles. We therefore introduce $\langle (\Delta R)^2\rangle$ to be the average displacement of hopping particles. We may now introduce softness directly and rewrite Eq.~\eqref{eq:diffusion_raw} as
\begin{equation}
    D \sim \langle (\Delta R)^2\rangle\int dS P(S) P_R(S).
\end{equation}
Finally, we make the mean-field approximation and write, $D\sim \langle (\Delta R)^2\rangle P_R(\langle S\rangle_{eq})$. As has been noted in previous studies~\cite{schoenholz2016structural}, $\langle(\Delta R)^2\rangle$ is only weakly temperature-dependent and so most of the temperature dependence should be contained in $P_R(\langle S\rangle_{eq})$.

Although we have significantly simplified the dynamics, we will see that this approximate framework suffices to explain the temperature dependence of silica. Combining the above arguments with the Arrhenius form of $P_R(S)$ we find,
\begin{align}
 \frac{1}{D(T)} &\propto \frac{1}{P_R(\langle S \rangle_{\mathrm{eq}}(T))}  \\
 & =\mathrm{exp}(\Delta E(\langle S \rangle_{\mathrm{eq}}(T))/T - \Sigma(\langle S \rangle_{\mathrm{eq}}(T))),
\label{eq:diffusivity}
\end{align}
where $\Sigma(S)$ and $\Delta E(S)$ are given by the fits shown in Fig~\ref{Fig2}(b), and $\langle S \rangle_{\mathrm{eq}}(T))$ is given by the fit in Fig~\ref{Fig3}(a). 

Since both $\Sigma(S)$ and $\Delta E(S)$ are approximately linear in the softness we can rewrite Eq.~\eqref{eq:diffusivity} as $\Sigma(S) = \Sigma_0 + \Sigma_1 S$ and $\Delta E(S) = E_0 + E_1 S$ respectively. As such, we can rewrite Eq.~\eqref{eq:diffusivity} as,
\begin{equation}
    \frac{1}{D(T)} \propto \exp\left[\left(\frac{E_0}T - \Sigma_0\right) - \left(\frac{E_1}T - \Sigma_1\right)S\right].\label{eq:diffs}
\end{equation}
Note that Eq.~\ref{eq:diffs} predicts that the diffusivity is independent of softness at the temperature $T_0=E_1/\Sigma_1$; this is the onset temperature (vertical dashed line in Fig.~\ref{Fig2}(a). Finally, since $S$ depends linearly on temperature we can write $S = S_0 + S_1 T$ and so the temperature dependence of Eq.~\eqref{eq:diffusivity} is given by
\begin{equation}
    \frac{1}{D(T)} \propto \exp\left[\frac{E_0 - E_1 S_0}{T} + \Sigma_1 S_1T\right].
\label{eq:proportionality}
\end{equation}
Thus we see that a crossover naturally emerges. When $\Sigma_1 S_1 T \gg (E_0 - E_1S) / T$ we expect silica to exhibit non-Arrhenius relaxation while at lower temperatures we recover strong liquid behavior, as expected.

\begin{figure}
\includegraphics[width=0.4\textwidth]{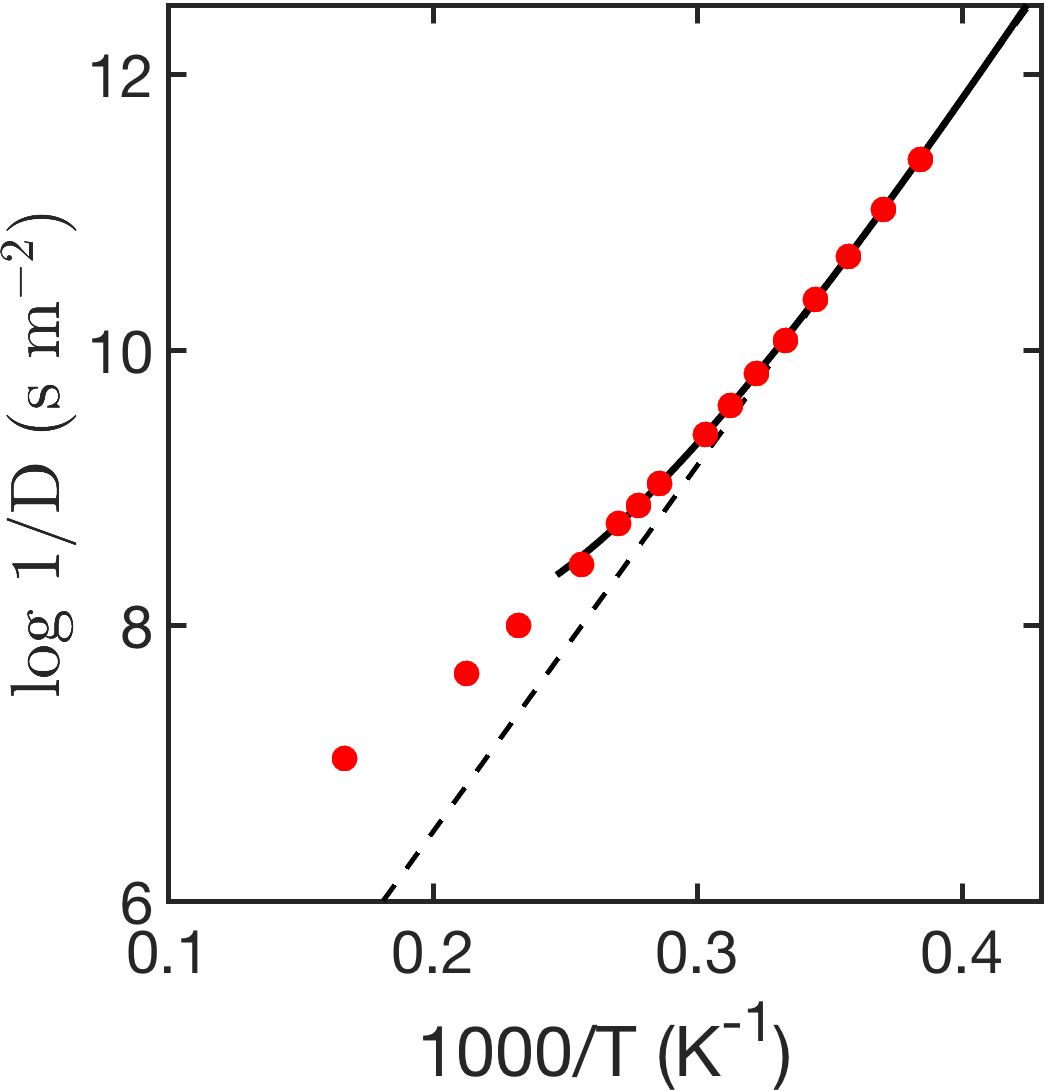}
\caption{
Inverse diffusivity vs. inverse temperature. Circles are measurements from MD simulations, and the solid line is our prediction based on \seq ~and the Arrhenius form. Note that the prediction is made only for temperatures below the onset temperature. The dashed line denotes the strong scaling, to reveal the strong-to-fragile crossover.  
}
\label{Fig4}
\centering
\end{figure}

We plot Eq.~\ref{eq:proportionality} in Fig.~\ref{Fig4} as a solid line. We note that the predicted $\frac{1}{D(T)}$ has a fragile-to-strong transition at approximately 3000 K. For a quantitative comparison, we also measure the inverse diffusivity of oxygen atoms in silica as a function of temperature, using the RMS displacement of the oxygen atoms, shown as red circles in Fig.~\ref{Fig4}. We used one fitting parameter to match the constant of proportionality in Eq.~\ref{eq:proportionality}. Evidently our prediction based on $\langle S \rangle_{\mathrm{eq}}(T))$  predicts the diffusivity of the atoms remarkably well over the entire temperature range studied, spanning the strong-to-fragile crossover. Note that the solid line only extends up to the onset temperature, as local structure is not relevant to dynamics above the onset temperature, by definition~\cite{schoenholz14}.        

In conclusion, we have shown that a simple model based on softness quantitatively predicts the temperature-dependence of relaxation in supercooled BKS silica. The same reasoning predicts the temperature-dependence of relaxation in a fragile glassformer, the Kob-Andersen model.  Note that although the method of calculating softness is the same for different systems, the actual definition of softness is based on the classification hyperplane, which depends on the system. Nevertheless, it has been shown that the emergent properties of softness in systems below yield exhibit commonality~\cite{cubuk2017structure}. Here we have shown that commonality in the emergent properties of softness extends even to the temperature-dependence of relaxation in glassformers of arbitrary fragility.  The fact that relaxation in both fragile and strong systems is well-described by the same simple reasoning in terms of average softness implies that softness provides a unifying framework for relaxation dynamics in glassy liquids. The difference in fragility arises from the differing dependences of average softness on temperature. For silica, the average softness is well-approximated as linear in $T$. For the Kob-Andersen system, the average softness is well-approximated by $\langle S \rangle = c_0 - c_1/T$. Our results shift the theoretical challenge from understanding fragility in glassforming liquids to understanding the temperature-dependence of a purely structural (static) quantity, $\langle S \rangle$.

Machine learning has already emerged as a promising tool for materials design~\cite{bassman2018active,sendek2020quantifying,cheon2018revealing,hoyt2019machine} as well as building conceptual models~\cite{hoffmann2019machine,rajak2019structural,rajak2019neural}. The ability to find predictive, compressed representations of physical data using machine learning becomes truly useful to theoretical physics if we can use those representations to build new models. This approach has the potential to be particularly useful for systems that are far out of equilibrium and/or disordered and/or that exhibit nonlinear response, where we cannot use statistical mechanics to bridge the gap between microscopic models and macroscopic behavior. In such situations, machine learning may provide a way of connecting microscopic information to collective behavior.

\section{Methods}

\subsection{Machine Learning Model}

As descriptors of local structure, we use structure functions that were used to predict the dynamics of Lennard-Jones particles and granular pillars~\cite{cubuk15}. These structure functions are closely related and are inspired by the symmetry functions that were proposed by Behler and Parrinello~\cite{behler07}. While these descriptors have been described in previous work in detail~\cite{cubuk15, schoenholz2016structural, schoenholz2017relationship, cubuk2017structure, sharp2018machine, ivancic2019identifying, ma2019heterogeneous, freitas2020uncovering}, we briefly introduce them here.  

Radial structure functions are given by,
\begin{align}
G(k;\mu) &= \sum_{i} \mathrm{e} ^{-(r_{ik}-\mu)^2/L^2}
\end{align}
which provides information about the radial density at distance $\mu$ from particle $k$, where $r_{ik}$ is the distance between particles $i$ and $k$. We choose the range of $\mu$ to be between 0 and 7 \AA. The parameter $L$ is the size of the window in radius, which is set to $L=0.2$~\AA. 

Angular structure functions are given by,
\begin{align}
\Psi(k;\xi,\lambda,\zeta) &= \sum_{i,j} \mathrm{e} ^{-\left(r_{ik}^2+r_{jk}^2+r_{ij}^2\right)/\xi^2}\left(1+\lambda \cos \theta_{kij} \right) ^\zeta
\end{align}
where $\xi$, $\lambda$, and $\zeta$ are variables that characterize angular structure functions; $\theta_{kij}$ is the angle made between particles $k$, $i$, and $j$.  These functions count the number of large and small bond angles within a distance $\xi$ of particle $k$. By varying $\lambda = \pm 1$, we can count large or small bond angles. $\zeta$ parameter controls the angular resolution of the structure functions. We identify rearrangements using the $p_{hop}$ measure, with a time window of 4 ps, and a cutoff of $p_{hop}=0.6s$.  The optimal C parameter of the linear SVM was found to be 1 by cross-validation. As mentioned in the main text, the training set and test set accuracies were both found to be 86\% when both radial and angular structure functions are used. When we restrict our model to only use radial structure functions, the prediction accuracy goes down to 79\%.

\subsection{Identifying Rearrangements}

To identify rearrangements we use the $p_{\text{hop}}$ metric that was first proposed by Candelier et. al.~\cite{candelier10} and has since been used extensively to identify rearrangements in amorphous materials~\cite{smessaert13, schoenholz2016structural}. To construct $p_{\text{hop}}$, first a timescale $t_R = 4$~ps is chosen to be commensurate with the timescale for rearrangements to take place in the system. Then two time intervals are defined as $A_t = [t - t_R / 2, t]$ and $B_t = [t, t + t_R / 2]$. For each particle $i$, $p_{\text{hop}}$ can be written as,
\begin{equation}
    p_{\text{hop}}(t) \sqrt{\langle (r_i - \langle r_i\rangle_B)^2\rangle_A\langle (r_i - \langle r_i\rangle_A)^2\rangle_B}
\end{equation}
where $\langle\rangle_A$ and $\langle\rangle_B$ are averages over the $A$ and $B$ interval respectively. When a particle is trapped in a cage, $\langle r_i\rangle_A \approx \langle r_i\rangle_B$ and $p_{\text{hop}}$ is equal to the scale of fluctuations about the cage center. However, when a particle has undergone a rearrangement, the means will shift and $p_{\text{hop}}$ will be exhibit a peak. The size of this peak will be commensurate with the size of the rearrangement.

\section{acknowledgments}
We would like to thank Austen Angell and Evan Reed for helpful discussions.

% \bibliography{bibliography.bib}

\end{document}